\documentstyle[12pt,epsfig,float,a4]{article}
\def\_{\rule{.3em}{.15ex}} 
\newcommand{\tb}  {\mbox{$ \tan\beta~ $}}

\newcommand{\be}{\begin{equation}}
\newcommand{\ee}{\end{equation}}
\newcommand{\bea}{\begin{eqnarray}}
\newcommand{\eea}{\end{eqnarray}}

\newcommand{\beq}{\begin{equation}}
\newcommand{\eeq}{\end{equation}}
\newcommand{\besg}{$b  \to  X_s \gamma~ $}
\newcommand{\mzero}{\rm m_0}
\newcommand{\mhalf}{\rm m_{1/2}}

\def\slash#1{\setbox0=\hbox{$#1$}#1\hskip-\wd0\dimen0=5pt\advance
       \dimen0 by-\ht0\advance\dimen0 by\dp0\lower0.5\dimen0\hbox
         to\wd0{\hss\sl/\/\hss}}
\def\gequiv{\raise 0.4ex \hbox{$>$} \kern -0.7 em \lower 0.62 ex \hbox{$\sim$}}
\def\gappeq{\mathrel{\rlap {\raise.5ex\hbox{$>$}}
{\lower.5ex\hbox{$\sim$}}}}

\begin{document}


\begin{flushright}
IEKP-KA/2001-03 \\[3mm]
{\tt hep-ph/0102163}
\end{flushright}


\begin{center}
  {\large\bf The   \besg  rate and  
          Higgs boson limits in the Constrained
Minimal Supersymmetric Model} \\[3mm]

  {\bf W. de Boer, M. Huber}
\\[2mm]
  {\it Institut f\"ur Experimentelle Kernphysik, University of Karlsruhe \\
       Postfach 6980, D-76128 Karlsruhe, Germany} \\[3mm]

  {\bf A.V. Gladyshev, D.I. Kazakov} \\[2mm]

{\it Bogoliubov Laboratory of Theoretical Physics,
Joint Institute for Nuclear Research, \\
141 980 Dubna, Moscow Region, Russian Federation}

\end{center}

\abstract{
New NLO \besg calculations
 have become available using resummed radiative corrections.
Using these calculations we perform a  global fit of the supergravity inspired
constrained minimal supersymmetric model.
We find that the resummed calculations show similar constraints
as the LO calculations, namely that only with a relatively heavy
supersymmetric mass spectrum   of $\cal{O}$(1 TeV) the $b-\tau$ Yukawa
unification and the \besg rate can coexist in the large $\tb$ scenario.
The resummed \besg calculations are found to
reduce the renormalization scale uncertainty considerably.
The low $\tb$ scenario is excluded by the present Higgs limits from LEP II.
The constraint from the Higgs limit in the $m_0,m_{1/2}$ plane
is severe, if the tri\-li\-near coupling $A_0$ at the GUT scale is fixed to
zero,
but is considerably reduced for $A_0\le -2m_0$. 
The relatively heavy SUSY spectrum required by \besg corresponds to a
Higgs mass of
$m_h=119\pm1~ \mbox{(stop masses)}~\pm2~(theory)~\pm3~\mbox{(top mass)}
~\rm GeV $ in the CMSSM. 
}


\section{Introduction}

In a previous paper we showed that the inclusive decay rate \besg \ at leading
order (LO) severely constrains the high $\tan\beta$ solution of the
Constrained Minimal 
Supersymmetric Standard Model (CMSSM)~\cite{PL}.
This was mainly caused by 
the fact that $b-\tau$ Yukawa coupling unification
preferred a negative sign 
for the Higgs mixing parameter $\mu$,
while the \besg \  rate required 
the opposite sign.
With initial
next-to-leading order (NLO)
calculations for the \besg \  rate in the MSSM \cite{ciu}
large terms proportional to $\tan^2\beta$
 changed the sign of $\mu$\cite{us}.
However, some sign errors in Ref. \cite{ciu}
were detected \cite{car,ciu1} and with the corrected sign we found that
no change of the sign of the chargino-stop amplitude occurred anymore, although
the NLO corrections were still large. 
Resumming of the large terms was done in Refs. \cite{car,ciu1}
and especially the contributions from the heavier stop, which were
neglected in Ref. \cite{ciu}, were taken into account.
Due to a large cancellation between the stop1 and stop2 contributions,
the NLO corrections turn out to be relatively small, so similar results as
in LO can be expected.
%

In our statistical  $\chi^2$ analysis the constraints from gauge
coupling unification, $b-\tau$ Yukawa coupling unification,
electroweak symmetry breaking,  \besg, relic density  and
experimental lower limits on SUSY masses can be considered
either separately or combined\cite{PL}.
Constraints from LO \besg rates were considered before in Refs.
\cite{bsg}.
As input parameters of the Constrained MSSM (CMSSM) we consider
 at the
unification scale (M$_{\rm GUT}$)
the unified gauge coupling constant ($\alpha_{\rm GUT}$),
the Yukawa coupling constants  of the third generation
(Y$_t^0$,Y$_b^0$,Y$_{\tau}^0$),
the common scalar mass (m$_0$),
 the common gaugino mass (m$_{1/2}$),
the common trilinear coupling (A$_t^0$=A$_b^0$=A$_{\tau}^0$),
and the Higgs mixing parameter $\mu^0$.
 Furthermore the ratio of the vacuum expectation
values of the two Higgs doublets (\tb) is a free parameter.
These parameters are optimized
  via a $\chi^2$ test to fit the low energy
experimental data on electroweak boson masses, \besg,
and quark and lepton masses of the third generation.
Details can be found in Refs. \cite{PL, ZP}.

 The values of m$_0$, m$_{1/2}$, $\mu^0$, A$^0$,  Y$^0$ and $\tan\beta$
 determine completely the mass spectrum of all SUSY particles
 via the RGE. The values of $\mu^0$, Y$^0$
 and $\tan\beta$ are constrained for given values of m$_0$ and
 m$_{1/2}$ by EWSB and the quark and lepton masses of the third generation.
 Since m$_0$ and m$_{1/2}$ are strongly correlated, we
 repeat each fit for every pair of m$_0$ and
 m$_{1/2}$ values between (200,200) and (1000,1000) GeV in steps of 100 GeV. 

\begin{figure}[htb]
\epsfig{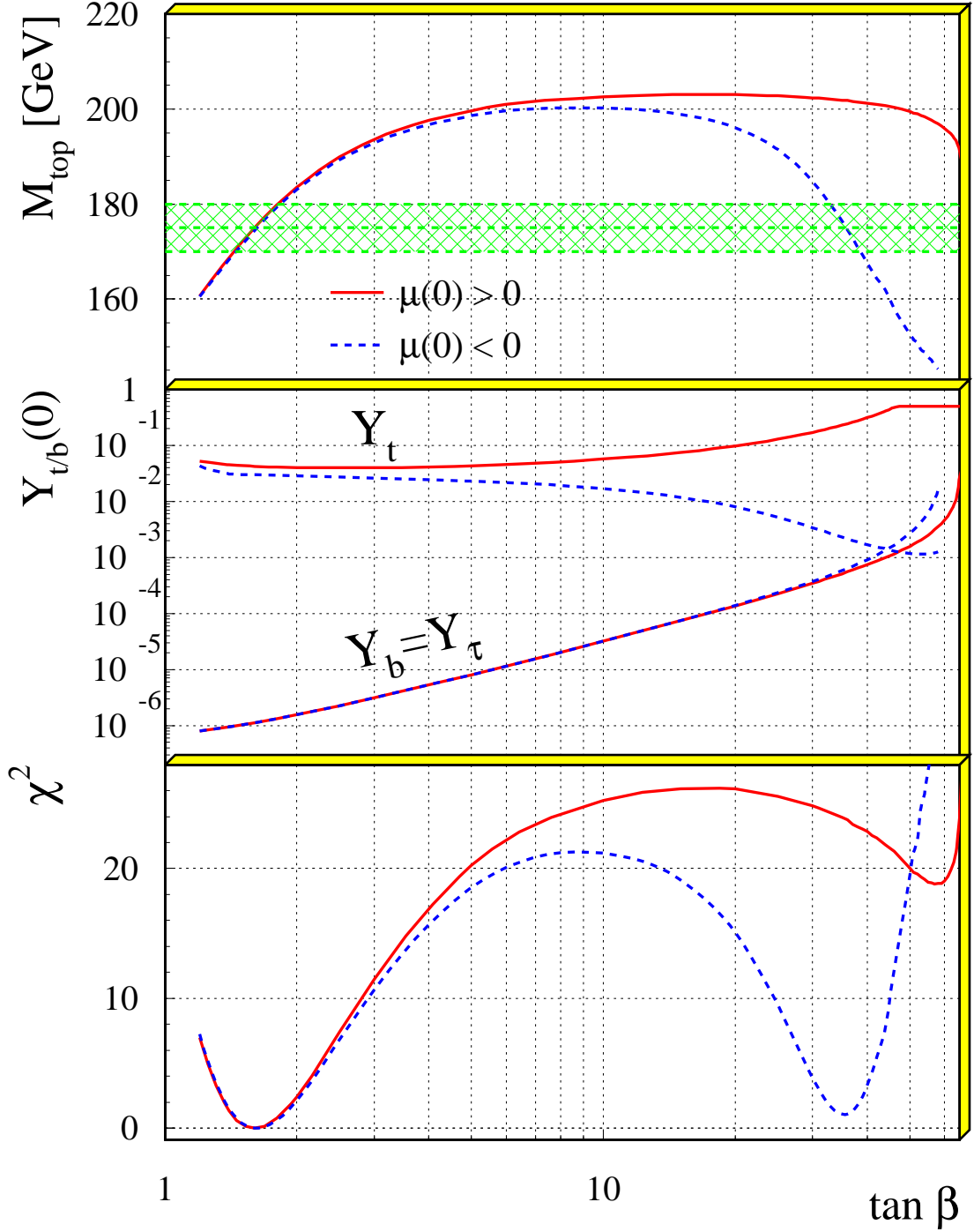}
\caption[]{\label{f1} 
The upper part shows the top quark mass as function of
tan $\beta$ for $\mzero$ = 600 GeV,$\mhalf$ = 400 GeV.
The middle part shows the corresponding values of the Yukawa
couplings at the GUT scale and the lower part the $\chi^2$ values, which
show that the value of tan $\beta$ are restricted to  the
following ranges 1$ < \tan\beta < $2 \  or 30$ < \tan\beta < $40
for $\mu < 0$.}
\end{figure}

The present value of the top mass of $174.3\pm 5.1$ \cite{pdb} constrains
$\tb$ to the regions $1.6\pm0.3$ and $35\pm3$ (see Fig. \ref{f1}),
which we call the low and high $\tb$ scenario, respectively.
The high $\tb$ scenario requires
 $\mu<0$, since the $\chi^2$ minimum of 19 for $\mu>0$ at
$\tb\approx 60$ is excluded now by the  top mass value.

At $\tb=35$ the top and bottom-tau Yukawa couplings are of the same order
of magnitude. Forcing  triple Yukawa unification would increase $\tb$ somewhat
(see middle part of Fig. \ref{f1}).  EWSB would then require
a slight splitting in the mass parameters of the two Higgs doublets,
which is possible by non-universal terms at the GUT scale\cite{split}.
However, the whole picture and mass spectra would not change significantly.


In the next section we consider NLO corrections to \besg in order to
check
whether they  are consistent with the $\mu<0$ solution at $\tb=35$,
since the solution at $\tb=1.65$ is excluded by the present Higgs limit
of 113.5 GeV\cite{newhiggs}, as will be discussed in the last section.

%

%
\begin{figure}[t]
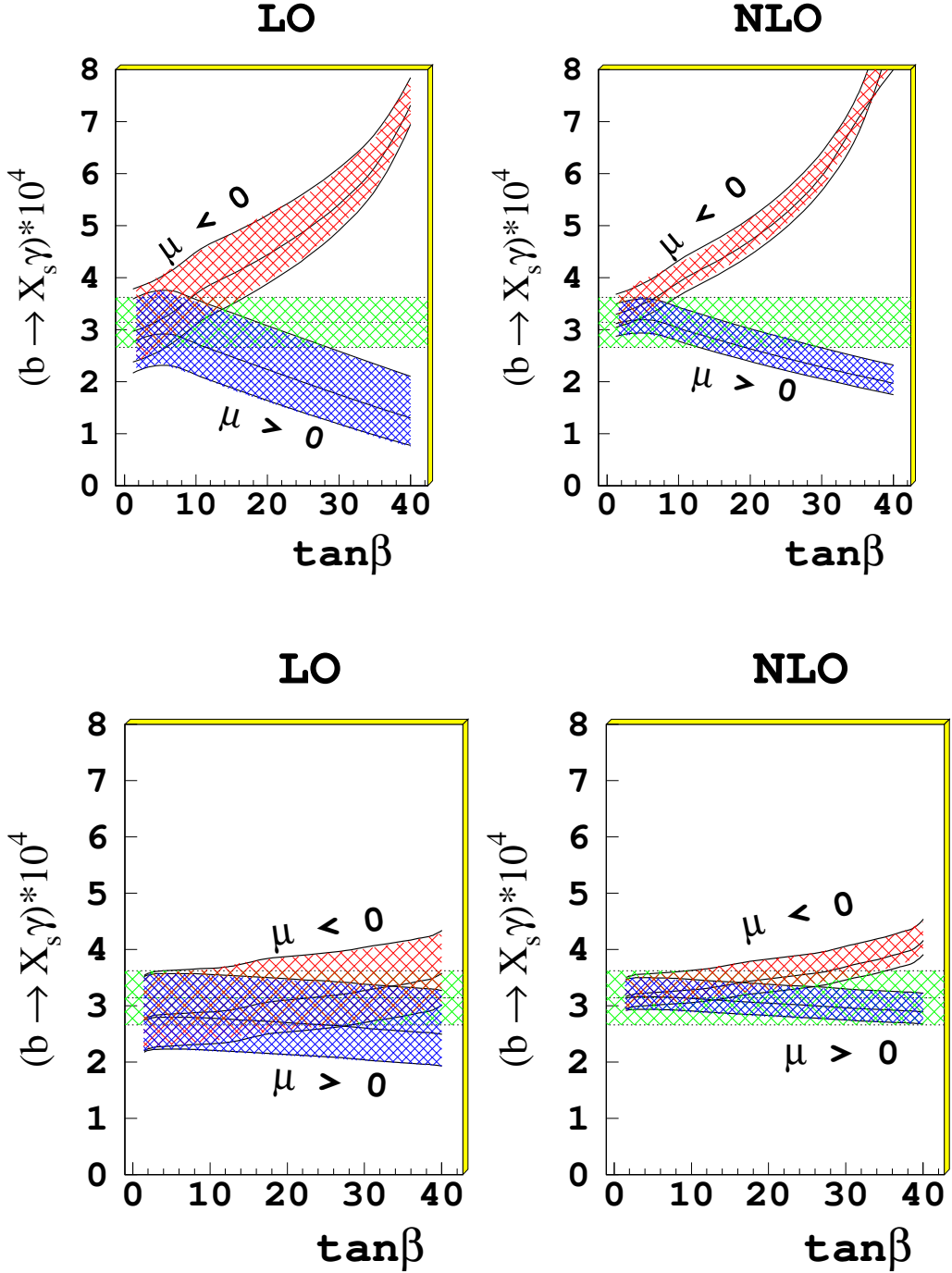
\vspace*{-4mm}
\epsfig{file=fig.2a,width=0.85\textwidth}  
\epsfig{file=fig.2b,width=0.92\textwidth}  
\caption[]{\label{f2} 
The dependence of the \besg \ rate on $\tan\beta$ for LO (l.h.s.)
and NLO (r.h.s.) for $A_0=0$ and   m$_0$ = 600 (1000) GeV, m$_{1/2}$ = 400
(1000)GeV at the top (bottom).
For each value of $\tan\beta$ a fit was made to bring the predicted \besg 
rate (curved bands) as close as possible to the data (horizontal bands).
The width of the predicted values shows
the renormalization scale uncertainty from a scale variation
between 0.5m$_b$ and 2m$_b$. Clearly, good agreement with the data
at large $\tb$ is only achieved for heavy sparticles.
}
\end{figure}
%
%
\begin{figure}[t]
\epsfig{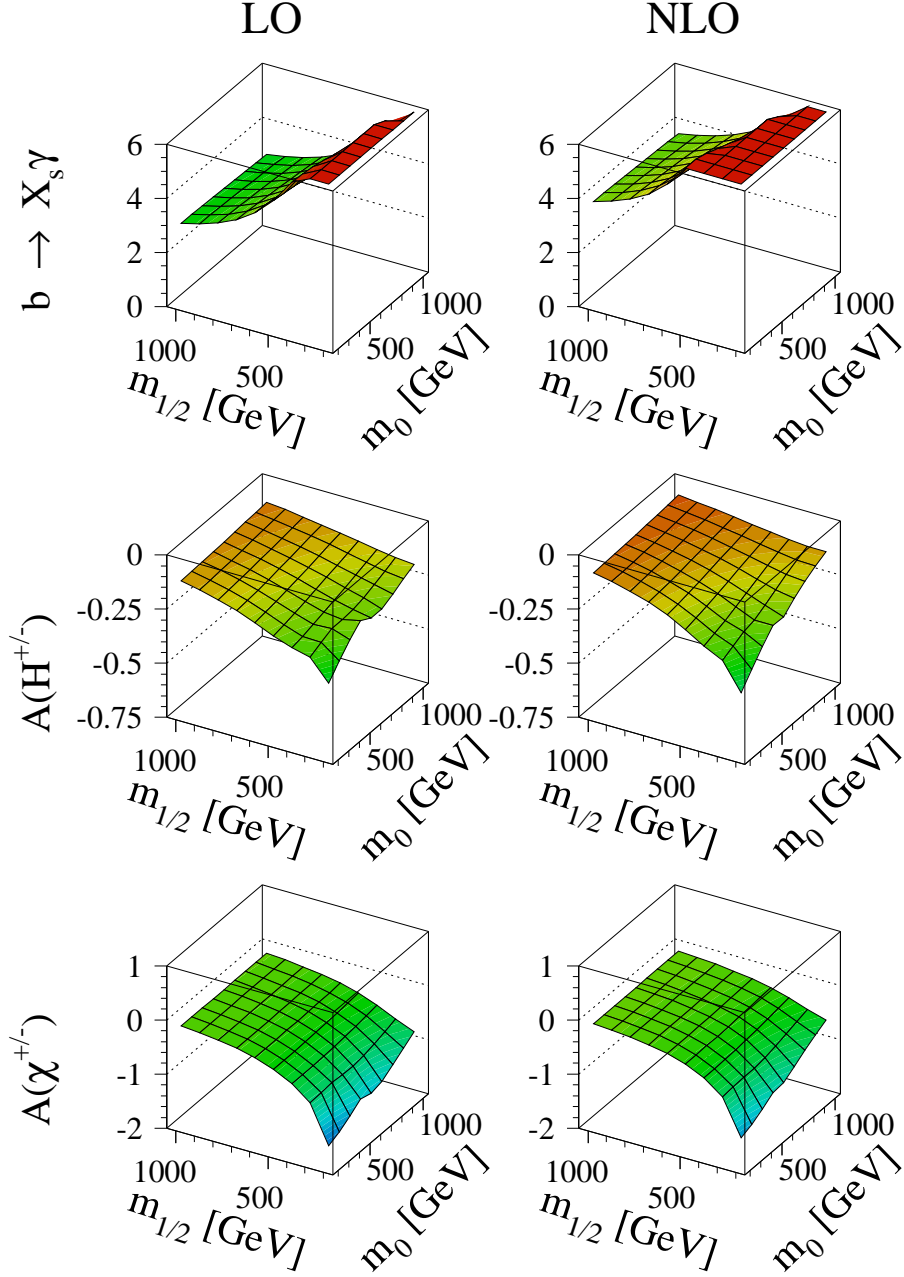}
\caption[]{\label{f3}
The decay rate (in units of $10^{-4}$) and selected amplitudes
(in units of $10^{-2}$) of the \besg \  decay for 
negative $\mu$ and $\tan\beta = 35$.
These amplitudes should  be  compared with the
SM amplitude of $-0.56\cdot 10^{-2}$.
}
\end{figure}
\begin{figure}[htp]
\begin{center}
\epsfig{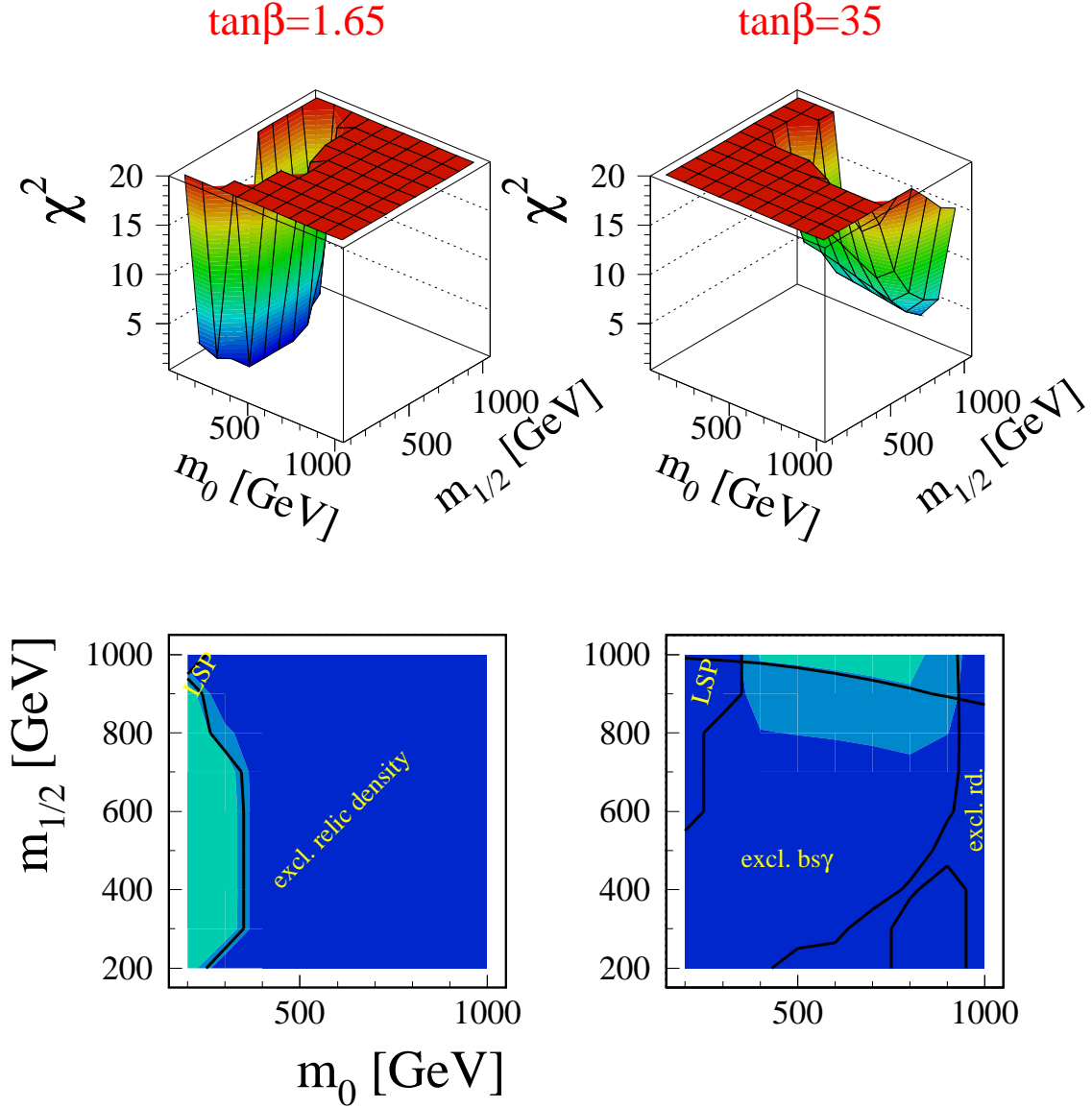}  
\caption[]{\label{f6}
The upper row shows the $\chi^2$ distribution in the $\mzero-\mhalf$ plane
for $\tan\beta = 1.65$ and $\tan\beta = 35$ with $\mu>0$ and $\mu<0$,
respectively.
The projections are shown in the second row.
The different shades in the projections
indicate steps of $\Delta\chi^2=4$. The contour lines show areas
excluded by the particular constraints used in the analysis:
in the LSP area the Lightest Supersymmetric Particle is charged
(usually the stau),
which is not allowed if the LSP is stable,
 in the relic density (rd) area the density of the universe is above
 the critical density and in the $bs\gamma$ area the \besg rate is too high.
The $\mu_b$ renomalization scale was allowed to vary
between 0.5m$_b$ and 2m$_b$ for the \besg rate.
}
\end{center}
\end{figure}


\section{ NLO  corrections to  \besg \ } 

We use  the \besg \
rate from the
CLEO Collaboration~\cite{stanf}:
$
BR(b \to X_s \gamma) = ( 3.15 \pm 0.35 \pm 0.32 \pm 0.26 ) \cdot 10^{-4}.
$
This value 
combined with the less precise ALEPH measurement~\cite{ALEPH} of 
$
BR(b \to X_s \gamma) = (3.11 \pm 0.80 \pm 0.72 ) \cdot 10^{-4}
$
yields as average
$ BR(b \to X_s \gamma) = ( 3.14 \pm 0.48 ) \cdot 10^{-4}
$

The \besg transition corresponds in lowest order
to a loop with either a W, charged Higgs or chargino. 
The leading order  corresponds to the emission of a
real photon from any of the charged lines, while the dominant
next-to-leading order (NLO) corrections involve virtual gluons
from any of the (s)quark lines.
The theoretical calculations of the \besg rate are well advanced.
The LO Standard Model (SM)
calculations~\cite{grin,BBMR,ajbur} have been complemented
by NLO calculations ~\cite{kadel,thurt1,aali,thurt2}.
Recently,  NLO calculations have been extended to
Two-Higgs Doublet Models (2HDM)~\cite{cdgg,borzu} and
the Minimal Supersymmetric Model (MSSM)~\cite{ciu,car,ciu1,mis}.
Here we use the results from Ref. \cite{ciu1}, which include all
potentially large two-loop contributions.


Fig. \ref{f2} shows the value of \besg decay rate as function of $\tb$
for two choices of the universal masses at the GUT scale, namely
$m_0=600, \ m_{1/2}=400$ and $m_0=1000, \ m_{1/2}=1000$ GeV.
In order to get
good agreement with the data at large $\tb$ one needs heavy sparticles,
as shown by the plots at the bottom.
The renormalization scale dependence in NLO is considerably
reduced as compared to the LO
calculations, as shown by the width of
the bands in
Fig. \ref{f2}. Here we only considered the dominant scale uncertainty
from the low energy scale, which was varied between 0.5$m_b$
and 2$m_b$.
The effect of the NLO calculations including both stops and
resummed corrections on the total rate and
individual amplitudes is rather small, as can be seen from the comparison
of the LO and NLO rate and am\-pli\-tudes shown in Fig. \ref{f3}.
So, the results at large $\tb$  are  similar 
to the previous LO calculations\cite{PL}, but
slightly more restrictive due to the smaller scale uncertainty in the NLO.

 Fig. \ref{f6} shows the $\chi^2$ distribution in the $\mzero,\mhalf$ plane
with the dominant source of the excluded regions  in the contours at
the bottom.
The analysis from Ref. \cite{olive}  finds a more restricted
region, mainly because
 they require the relic density to be between 0.1 and 0.3, while
we require  simply that the relic density is not above the critical density.
The low $\tb$ scenario from Ref. \cite{PL} is shown for comparison.
Previously it was constrained mostly by the relic density,
although now the whole low \tb scenario is excluded by the present
limits on the Higgs mass from LEP,
as will be discussed in the next
section.



\section{Higgs mass constraints}

In Supersymmetry the couplings in the Higgs potential
are the gauge couplings. The absence of arbitrary couplings
together with well defined radiative corrections to the masses
results in clear predictions for the lightest Higgs mass
and electroweak symmetry breaking (EWSB)\cite{rev}.

In the Born approximation one expects the lightest Higgs to have a
mass $m_h$ below the $Z^0$ mass. However, loop corrections,
especially from top and stop quarks, can increase $m_h$
considerably\cite{radhiggs}. 
The Higgs mass depends mainly on the following parameters: the top mass,
the squark masses, the mixing in the stop sector,
the pseudoscalar Higgs mass and $\tan\beta$. 
As will be shown below, the maximum Higgs mass is obtained for 
large $\tan\beta$, for a maximum value of the top and squark masses and 
a minimum value of the stop mixing.
The Higgs mass calculations were carried out following the results
obtained by 
Carena, Quir\'os and Wagner\cite{carenawagner} in a renormalization 
group improved effective potential approach,  including the 
dominant two-loop contributions from gluons and gluinos.
The gluino contributions were taken from the FeynHiggs
calculations\cite{feynhiggs}.

Note that in the CMSSM the Higgs mixing parameter $\mu$ is determined 
by the requirement of EWSB, which yields large values for $\mu$\cite{rev}. 
Given that the pseudoscalar Higgs mass increases rapidly with $\mu$,
this mass is always much larger than the lightest Higgs mass and
thus decouples. 
We found that this decoupling is effective
for all regions of the CMSSM parameter space, i.e. the lightest Higgs 
has the couplings of the SM Higgs within a few percent. 
Consequently, the experimental  limits on the SM Higgs can be taken.

\begin{figure}[htb]
\begin{centering}
\epsfig{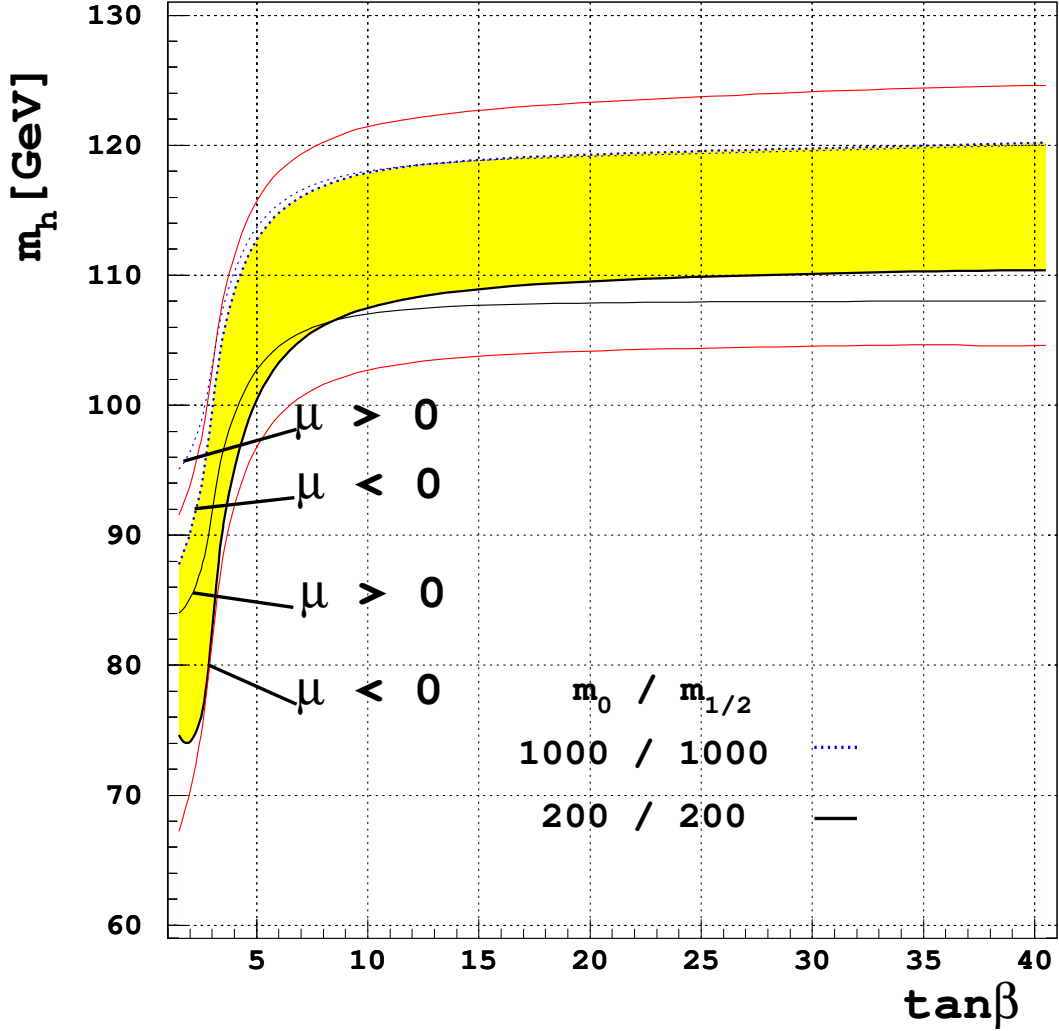}  
\caption[The mass of the lightest  Higgs boson]
{\label{mhiggstanb}
The mass of the lightest Higgs boson as
function of $\tan\beta$,
as calculated by the effective potential
approach\cite{carenawagner}.
The shaded band shows the variation of $\mzero=\mhalf$ between
200 and 1000 GeV  for $\mu<0$, $m_t=175$ GeV, and $A_0=0$.
Note the small dependence
on the sign of $\mu$ for large $\tb$, as expected from the
suppression of $\mu$ by \tb in the stop mixing.
The maximum (minimum) Higgs boson mass value, shown by the upper (lower) line
are obtained for $A_0=-3\mzero$, $m_t=180$ GeV, $\mzero=\mhalf=1000$ GeV
($A_0=3\mzero$, $m_t=170$ GeV, $\mzero=\mhalf=200$ GeV).
As can be seen the curves show an asymptotic behaviour for large values
of $\tan\beta$.
} \end{centering} \end{figure}

The lightest Higgs boson mass $m_h$ is
shown as function of $\tan\beta$  in
Fig.~\ref{mhiggstanb}. The shaded band corresponds
to the uncertainty from the stop mass  for $m_t=175$ GeV.
The upper and lower lines correspond to $m_t$=170 and 180 GeV, respectively. 

One observes that for a SM Higgs limit of 113.5 GeV \cite{newhiggs} all
values of \tb below 4.3 are excluded in the CMSSM.
This implies that the low $\tb$ scenario with $\tb=1.6\pm 0.3$
(see Fig. \ref{f1})
is excluded.

\begin{figure}[t]
\epsfig{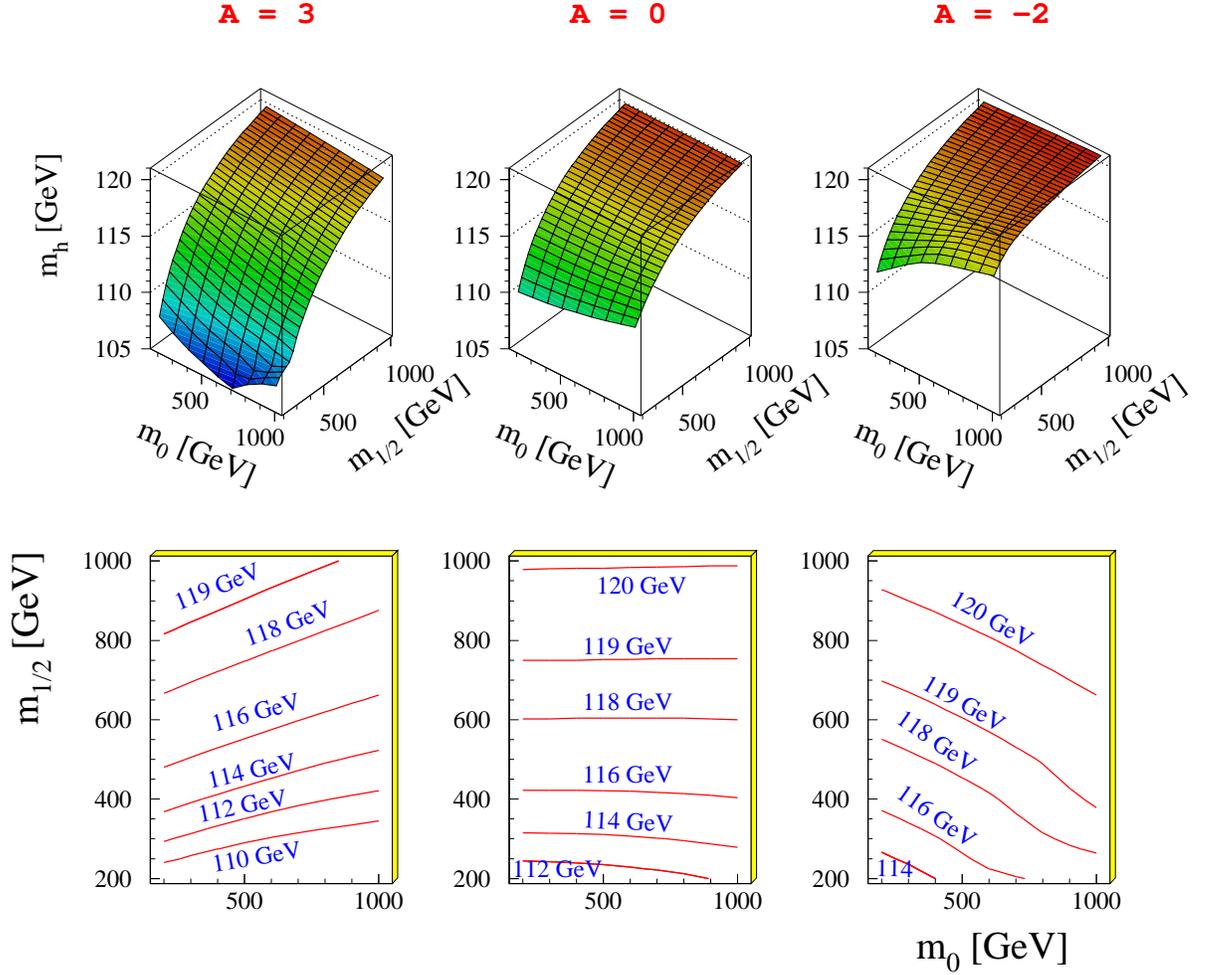}
\caption[]{\label{hi35}
The Higgs boson mass  as function of $\mzero$ and $\mhalf$,
for three values of the trilinear coupling $A$ in units of $\mzero$
as calculated by the effective potential approach\cite{carenawagner}. 
 The top mass is 175 GeV and $\tb$ is large, so the results
are independent of the precise value of $\tb$. Cleraly, for $A=-2\mzero$
the excluded area by the present Higgs limit of 113.5 GeV is only the
small lefthanded corner.
}
\end{figure}

In order to estimate  the uncertainties of the
Higgs mass predictions in the CMSSM,
 the relevant parameters were varied one by one.
The Higgs mass varies between 110 and 120 GeV, if $\mzero$ and $\mhalf$
are varied between 200 and 1000 GeV, which implies stop masses
varying between 400 and 2000 GeV, as shown in Fig. \ref{hi35}
for three different values of the trilinear coupling  at the
GUT scale $A_0$ in units of $\mzero$. For larger values of $\mhalf$ and
$\tb$
the Higgs mass saturates, as is obvious from the 3-D plots in Fig. \ref{hi35}
and from Fig. \ref{mhiggstanb}.
The \besg rate requires
$A_0 > -2m_0$, since lower values increase \besg even more for $\mu<0$.
For $A_0=-2m_0$   $m_h>113.5$ GeV
excludes  only the small corner at the left bottom in
the $\mzero, \mhalf$ plane .
In  Ref. \cite{olive} the excluded region is larger,
since they kept $A_0=0$.
Note that the stop mixing parameter $X_t=A_t-\mu/\tb$ is not an arbitrary
free parameter in the CMSSM, since the Higgs mixing parameter $\mu$ is
determined
by EWSB and the value of $A_t$ at low energy is largely determined by
 $\mhalf$ through radiative corrections, so the Higgs mass uncertainty
from the stop mixing is included in the variation of 
$\mzero$ and $\mhalf$. One observes from Fig. \ref{mhiggstanb}
that at large $\tb$ the Higgs mass varies
between 110 and 120 GeV, if the SUSY mass parameters are varied up to 1
TeV. We take the variance of this interval, which is 10/$\sqrt{12}$=3 GeV,
as an error estimate for the uncertainty from the stop masses.
The values  $\mzero=\mhalf=370$ GeV 
yield the central value of $m_h=115$ GeV.
%

\begin{figure}[htb]
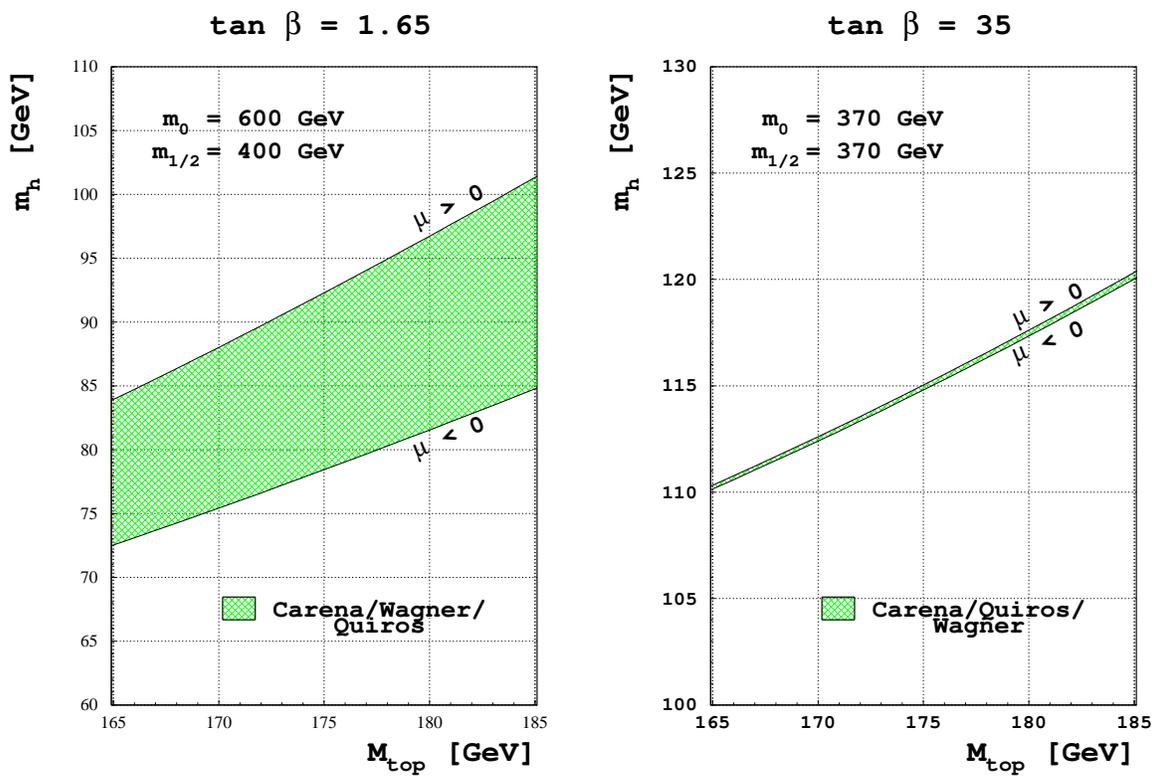

\begin{centering}
\epsfig{file=fig.7a,width=.49\textwidth}  
\epsfig{file=fig.7b,width=.49\textwidth}  
\caption[Dependence of the Higgs mass on $M_{\rm top}$]{\label{mtopmh}
The top mass dependence of the Higgs mass  in the low and high $\tb$ scenario.
Note the reduced dependence on the sign of $\mu$ for large \tb.
}\end{centering}
\end{figure}

The dependence on $m_t$ is shown in Fig. \ref{mtopmh} for $A_0=0$ and 
intermediate values of $\mzero$ and $\mhalf$ for two values of $\tb$ 
(corresponding to the minimum $\chi^2$ values in Fig. \ref{f1}). 
The uncertainty from the 
top mass at large $\tb$ is  $\pm$ 5 GeV, given the
uncertainty on the top mass of 5.2  GeV.
The uncertainty from the higher order calculations (HO) is estimated to be
2 GeV from a comparison of the full diagrammatic method \cite{feynhiggs}
and the effective potential approach\cite{carenawagner}, 
so combining all the uncertainties discussed before we find for
the  Higgs mass in the CMSSM
\be
m_h=115\pm3~ \mbox{(stop masses)}~\pm2~(theory)~\pm5~\mbox{(top mass)}
~\rm GeV, \label{mh} 
\ee
where the errors  are the estimated standard deviations around the central
value.
As can be seen from Fig.~\ref{mhiggstanb} this  central value 
is valid for all $\tan\beta > 20$ and decreases for lower $\tan\beta$.

If we include previous constraints from \besg, than  the allowed region
is restricted by $m_{1/2}>700$ GeV, leading to heavy stop masses. This
results in a heavier Higgs mass with a reduced error due to the saturation
of the Higgs mass for large sparticle masses: 
\be
m_h=119\pm1~ \mbox{(stop masses)}~\pm2~(theory)~\pm3~\mbox{(top mass)}
~\rm GeV.  
\ee
\section{Conclusions}

The results can be summarized as follows:
\begin{itemize}
\item
The NLO \besg contributions do not strongly change the  LO predictions; only
the renormalization scale uncertainty decreases, thus
increasing the excluded parameter region. The observed \besg is still
difficult to reconcile at large $\tb$ with $b-\tau$ unification,
as observed before in LO\cite{PL}.
\item
The low $\tb$ scenario ($\tb<4.3$) of the CMSSM is excluded by the
95 \% C.L.  lower limit on the Higgs mass of  113.5 GeV\cite{newhiggs}.
\item
For the high $\tb$ scenario the  Higgs mass is found to be below 125 GeV
in the CMSSM.
This prediction is independent of $\tb$ for $\tb>20$
and decreases for lower $\tb$.
The Higgs mass corresponding to the relatively heavy sparticle spectrum
required by the \besg measurement is:
$m_h=119\pm1~ \mbox{(stop masses)}~\pm2~(theory)~\pm3~\mbox{(top mass)}
~\rm GeV.$  
\item
The constraint  in the $\mzero,\mhalf$ plane by the present Higgs limit
of 113.5 GeV is severe, if the trilinear coupling $A_0$ is choosen to be
positiv at the GUT scale, but is strongly reduced  for $A_0\le -2m_0$.
\end{itemize}
\section*{Acknowledgements}
A.G. and D.K. would like to thank the Heisenberg-Landau Programme
and RFBR grant \# 99-02-16650 for financial support
and  the Karlsruhe University for hospitality during the time this work
was carried out.
\section*{Note added in proof.}
Recently the Muon (g-2) Collaboration (H.N. Brown et al., 
hep-ex/0102017) has measured the anomalous magnetic moment of the 
muon. The find a 2.6$\sigma$ deviation
from the SM expectation. If interpreted as 
contributions from supersymmetry, then this measurement would 
require $\mu>0$ in our conventions. This would be opposite to
the sign required by $b-\tau$ unification, so to include this measurement
would require to give up $b-\tau$ unifcation or to modify the CMSSM, e.g.
by including complex phases. This will be studied in a forthcoming paper.

%
%
%
%
\end{document}